\documentclass[aps,prb,twocolumn,amsmath,amssymb]{revtex4}
\usepackage{graphicx}
\usepackage{bm}
\usepackage{color}
\usepackage{psfrag}
\usepackage{multirow}%
\usepackage{times}%

\newcommand{\diag}[2][]{\parbox[c]{#2\linewidth}{\includegraphics*[width=\linewidth]{#1}}}
\newcommand{\Addsp}[2][]{\parbox[c]{#2pt}{\rule{#2pt}{#1}}}

\begin{document}
\title{A Generalized Quantum Dimer Model Applied to the Frustrated Heisenberg Model on the Square Lattice: Emergence of a Mixed Columnar-Plaquette Phase}
\author{A. Ralko,${^1}$ 
M. Mambrini,${^2}$ and D. Poilblanc$^{2}$ }
\affiliation{
${^1}$ Institut N\'eel, CNRS 
and Universit\'e de Grenoble, F-38000 France \\
${^2}$ Laboratoire de Physique Th\'eorique, CNRS 
and Universit\'e de Toulouse, F-31062 France
} 
\date{\today}
\begin{abstract}
Aiming to describe frustrated quantum magnets with non-magnetic singlet ground states,
we have extended the Rokhsar-Kivelson (RK) loop-expansion  to derive a generalized 
Quantum Dimer Model  containing only connected terms up to arbitrary order.
 For the square lattice frustrated Heisenberg antiferromagnet ($J_1$-$J_2$-$J_3$ model), an expansion up to 8th order
 shows that the leading correction to the original RK model comes from dimer moves on length-6 loops.
 This model free of the original sign problem is treated by advanced numerical techniques. The results suggest that a
rotationally anisotropic plaquette phase [\onlinecite{ralko1}]
 is the ground state of the Heisenberg model in the parameter region of maximum frustration.
\end{abstract}
\maketitle

Over the last decades theoretical efforts have been devoted to
study new quantum phases of bidimensional frustrated quantum magnets,
motivated by the discovery of experimental antiferromagnets showing the
absence of long-range
magnetic ordering down to very low
temperatures.\cite{herbertsmithite,chaboussant,shastry,linio2,robert} 
In such sytems a gap to magnetic excitation traditionally opens up
while the spin-SU(2) symmetry remains unbroken. Two classes of ``singlet'' phases 
have been distinguished: Valence Bond Crystals (VBC) where some spatial symmetries
are spontaneously broken, and Spin Liquid (SL) for which all symmetries remain
unbroken ({\it e.g.} the resonating valence bond (RVB)
liquid.\cite{anderson}).

However, it is usually difficult to characterize the singlet phases in simple microscopic $S=1/2$ models. 
For example, in the well-known 
$J_1-J_2$ Heisenberg $S=1/2$ antiferromagnet on the square lattice, where
frustration is controlled by the next-nearest interaction $J_2$,  no definitive answer has been
given on the nature of the non-magnetic GS for maximal frustration at
$J_2/J_1\sim0.5$.
M.~Mambrini {\it et al.} recently addressed a work to this task \cite{mambrini},
studying the $J_1-J_2-J_3$ model, containing an extra next-next-nearest neighbor
$J_3$ frustrating antiferromagnetic,
\begin{eqnarray}
	\label{eq:heisenberg}
	{\cal H} = J_1 \sum_{\langle i,j\rangle} {\bf S}_i . {\bf S}_j + J_2
\sum_{\langle \langle i,j\rangle \rangle} {\bf S}_i . {\bf S}_j  + J_3
\sum_{\langle \langle \langle i,j\rangle \rangle \rangle} {\bf S}_i . {\bf
S}_j.
\end{eqnarray}
Interestingly, in this model the description in terms of
Nearest Neighbor Valence Bond (NN VB) coverings~\cite{note_VB} is excellent in some extended region of parameter space, in particular 
around the point $J_2=J_3=1/4$, with some significant extension along the line $(J_2+J_3) = J_1 / 2$.
This model is therefore one simple canonical case where a truncation within the nearest-neighbor singlet 
configuration basis is legitimate and can be used as a simpler and convenient framework.

In this paper, we have extended the Rokhsar-Kivelson (RK) loop-expansion  to derive a generalized 
Quantum Dimer Model (QDM) {\it acting in the space of
hardcore NN dimer coverings of the lattice}. We show that this expansion based on the hierarchy of the
overlap matrix elements between the dimer coverings  leads to an effective Hamiltonian 
that contains a sum of dimer moves, each involving only a single closed loop or loops at finite distances (connected term). 
In other words, all disconnected terms cancel out order by order.
We apply this procedure to the $J_1-J_2-J_3$ model and show that the leading contributions are of the
form of a simple generalized QDM on the square lattice which, in addition to
original QDM \cite{rokhsar},
contains an additional loop-6 term which brings kinetic
competition in the system. The effective Hamiltonian then reads:
\begin{eqnarray}
\label{eq:gqdm}
 {\cal H}^{\textrm{eff}} = v \sum_{p} \diag[g_1_2]{0.06}_{p}
- t_4 \sum_{p} \diag[g_1_1]{0.06}_{p}  - t_6 \sum_{p} \diag[g_2_2]{0.08}_{p}.
\end{eqnarray}
where the sums run respectivelly over all square or rectangular plaquettes of the square lattice. Here we use the following
convention: (i) White plaquettes denote {\em kinetic} (off-diagonal) operators
that flip dimers around the thick contour (ii) Yellow plaquettes stand for
{\em potential} (diagonal) operators that leave configurations unchanged with
a factor $1$ if it is flippable around the thick contour and $0$ in the opposite case. In the following, the ommision of $p$ indices in the diagrammatic notation is a shortcut notation for an implicit summation over all inequivalent plaquettes with a given shape. For example, $\diag[g_1_2]{0.06} = \sum_{p} \diag[g_1_2]{0.06}_{p}$.

Guided by a variational
approach and combining numerical techniques such as Exact Diagonalizations (ED) and
Zero-temperature Green function Monte-Carlo (GFMC), we compute the phase
diagram of this model. More specifically, we give evidence in favour of a
Mixed Columnar-Plaquette phase first proposed in~[\onlinecite{ralko1}] and,
since, evidenced in number of other contexts~\cite{trousselet}.
Remarkably, this phase is found to be stable even in the presence of the loop-6 fluctuations. 
Hereafter, using the relation between the effective and microscopic models, 
we argue  in favor of  the 
SU(2)-invariant version (i.e. applicable to a spin-1/2 model instead of a dimer model) of the above Mixed Columnar-Plaquette phase in the  $J_1-J_2-J_3$ microscopic model
along the maximally frustrated line $J_2+J_3=J_1/2$ where an approach restricting to the
short-range VB basis has been justified previously \cite{mambrini}.

{\it Derivation of the model}. A systematic way to derive the generalized QDM hamiltonian (\ref{eq:heisenberg}) consists in (i) Projecting the Heisenberg model in
the manifold formed by NN VB coverings of the square lattice, (ii) Perform the unitary transformation
that turns the non-orthogonal VB basis into the orthogonal Quantum Dimer basis. The key ingredient of the calculation is the overlap matrix ${\cal O}_{\varphi,\psi}=\langle \varphi | \psi \rangle$ where $|\varphi\rangle$ and $| \psi \rangle$ are two NN VB states.
Step (i) is equivalent to solving the generalized eigenvalue problem ${\cal H} | \varphi \rangle= E {\cal O} | \varphi \rangle$,
while the orthogonalization required in step (ii) is conveniently achieved by defining ${\cal H}^{\text{eff}} = {\cal O}^{-1/2} {\cal H} {\cal O}^{-1/2}$.

Using a convention where all bond singlets are oriented from sites A to sites B according to the canonical bipartition of the square lattice, the overlap matrix can be written as ${\cal O}_{\varphi,\psi}= \alpha^{N - 2n_l(\varphi,\psi)}$
where $N$ is the size of the system, $n_l(\varphi,\psi)$ the number of loops in the ovelap diagram obtained by superimposing the two configurations, and $\alpha=1/\sqrt{2}$. On the other hand, $\langle \varphi | {\bf S}_i . {\bf S}_j | \psi \rangle= \varepsilon \langle \varphi | \psi \rangle$ with $\varepsilon=-3/4$ (resp. $\varepsilon=+3/4$) if $i$ and $j$ belongs to the same loop of the overlap diagram but belong to distinct sublattices
(resp. belong to the same sublattice) and $\varepsilon=0$ if $i$ and $j$ belongs to two distinct loops. Using a convenient scaling and shifting ${\cal H} \rightarrow (4/3){\cal H}+ J_1 N /2$ of the Hamiltonian (\ref{eq:heisenberg}), the matrix elements $\langle \varphi | {\cal H} | \psi \rangle$
can be expressed as ${\cal H}_{\varphi,\psi}= h_{\varphi,\psi} {\cal O}_{\varphi,\psi}$ where $h_{\varphi,\psi}$ only depends of the loops configuration. In particular, this convention enforces $h_{\varphi,\varphi}=0$ for all $\varphi$. 

It is then possible to expand ${\cal O}$ and ${\cal H}$ in powers of $\alpha$ and compute ${\cal H}^{\text{eff}} = {\cal O}^{-1/2} {\cal H} {\cal O}^{-1/2}$
accordingly as shown in table (\ref{tab:expansion}) up to $\alpha^6$. The
expansion up to $\alpha^8$ as well further technical details of the
calculation are given in the appendix \ref{Derivation}. It is worth
mentionning two peculiarities of this expansion : (i) contrary to several
previous approaches~\cite{rokhsar,ralko2,moessner} our expansion is {\it not} 
controlled by the {\em length of the loops}, but by the actual amplitudes of the overlap matrix elements that only depend on the {\em overall number of loops} in the overlap diagrams, (ii) all non-local and disconnected processes appearing in both ${\cal H}$ and ${\cal O}$ cancel in the expression of ${\cal H}^{\text{eff}}$.

\begin{table}
\begin{center}
\begin{tabular}{|c|c|c|c|} \hline\hline
 Processes 				  								& ${\cal O}$ 	& ${\cal H}$ 										&	${\cal H}^{\textrm{eff}}={\cal O}^{-1/2} {\cal H} {\cal O}^{-1/2}$ \\ \hline
 Id 																& 1						&  0														& 0																															\\ \hline
 \Addsp[6mm]{0.} \diag[g_1_2]{0.06} & $\emptyset$	&  $\emptyset$									& $2(J_1-J_2)\alpha^4$ 																					\\ \hline
 \Addsp[6mm]{0.} \diag[g_1_1]{0.06} & $\alpha^2$	&  $2(-J_1+J_2)\alpha^2$ 				& $-2 \left(J_1-J_2\right) \alpha ^2 \left(1+\alpha ^4\right)$	\\ \hline
 \Addsp[6mm]{0.} \diag[g_2_2]{0.08}	& $\alpha^4$	&  $2(-2J_1+2J_2+J_3)\alpha^4$	& $2 \left(-J_1+J_2+J_3\right) \alpha ^4$												\\ \hline
 \Addsp[6mm]{0.} \diag[g_2_1]{0.10} & $\alpha^4$	&  $4(-J_1+J_2)\alpha^4$ 				& 0																															\\ \hline
 \Addsp[6mm]{0.} \diag[g_3_2]{0.12} & $\alpha^6$	&  $2(-3J_1+3 J_2+J_3)\alpha^6$	& 0																															\\ \hline
 \Addsp[8mm]{0.} \diag[g_3_3]{0.08} & $\alpha^6$	&  $2(-3J_1+3 J_2+2J_3)\alpha^6$&	$2 \left(-J_1+J_2+J_3\right) \alpha ^6$												\\ \hline
 \Addsp[6mm]{0.} \diag[g_3_4]{0.10} & $\alpha^6$	&  $2(-3J_1+3 J_2+2J_3)\alpha^6$&	$\left(-J_1+J_2+2 J_3\right) \alpha ^6$												\\ \hline
 \Addsp[6mm]{0.} \diag[g_3_1]{0.14} & $\alpha^6$	&  $2(-J_1+J_2)\alpha^6$				& 0																															\\ \hline
 \Addsp[6mm]{0.} \diag[g_2_4]{0.08} & $\emptyset$	&  $\emptyset$									& $\left(J_1-J_2-J_3\right) \alpha ^6$													\\ \hline\hline
\end{tabular}
\end{center}
\caption{\label{tab:expansion} Expansion of ${\cal O}$, ${\cal H}$  and ${\cal
H}^{\textrm{eff}}$ up to order $\alpha^6$. Note that the effective hamiltonian
contains only local (connected) processes. Some processes (marked as
$\emptyset$) does not appear in ${\cal O}$ or ${\cal H}$, but are produced in
${\cal H}^{\textrm{eff}}$ by contractions of the generically non-commuting
terms of the expansion (see appendix \ref{Derivation} for details). }
\end{table}

Let us discuss the results of this expansion. When truncated up to order
$\alpha^2$, we recover the usual hamiltonian obtained in [\onlinecite{rokhsar}] with $v/|t_4|=\alpha^2=1/2$. Note that such a drastic truncation appears a bit pathological in the sens that it does not capture any aspect of the frustration of the original model : $v/|t_4|$ is indeed independant of $J_2/J_1$. In the perspective of a justification of QDM model from the Heisenberg model, non trivial effects
emerge from order $\alpha^4$. Furthermore, considering the last column of table (\ref{tab:expansion}) it is quite easy to see that, in the maximally frustrated region of the phase diagram ($J_2+J_3 \sim J_1/2$) where the validity of the NN VB approach have been established~\cite{mambrini}, only the 3 processes retained
in (\ref{eq:gqdm}) are dominant. Importantly, we find that $t_4>0$ and $t_6>0$ which enable the use of efficient stochastic methods not applicable to the original frustrated spin model
which suffers from the so called ``minus sign'' problem.



\begin{figure}
\includegraphics[width=0.40\textwidth,clip]{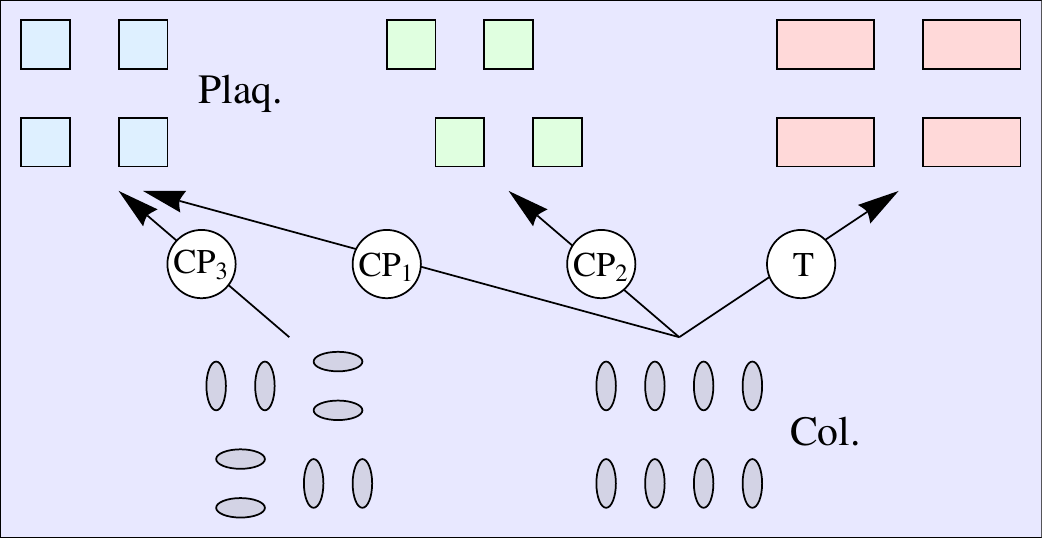}
\caption{\label{fig01} (Color online). VBC states considered in this work.  
Generalized anisotropic VBC states labeled by $CP_a$ (mixed columnar-plaquette phases) and $T$ 
(trimerized phase) interpolate between the most symmetric limits shown on the figure.}
\end{figure}

{\it Variational analysis:} We now turn to the investigation of the effective Hamiltonian (\ref{eq:gqdm}).
We start with some
discussion of the expected VBC phases shown on Fig.~\ref{fig01}.
Regular columnar and plaquette phases have been introduced in the 
context of the frustrated $J_1-J_2$ model and of the QDM \cite{j1j2}. More recently,
an anisotropic mixed columnar-plaquette phase has been introduced \cite{ralko1}.
We consider here the possibility of such phases which interpolate 
between the simple higher-symmetry VBC (such as columnar or plaquette).
Because of the presence of loop-6 dimer moves, we also consider the possibility
of a trimerization of the columns of dimers. 
We summarize the quantum numbers of the degenerate GS 
of the various VBC in table~\ref{tab:symmetries}. This will
be used further in this paper to analyze the low-energy spectrum of Hamiltonian
(\ref{eq:gqdm}).

\begin{table}
\begin{center}
\begin{ruledtabular}
\begin{tabular}{|*{9}{c|}}
&$\Gamma,A_1$&$\Gamma,B_1$&$M,A_1$&$K,A_1$&$K,B_1$&$Q_B$ &$Q_2$& $Q_3$\\
\hline
{\rm Col.}   &$\surd$&$\surd$&$\surd$  & & & & & \\
{\rm Pla.}    &$\surd$& &$\surd$ &$\surd$ & & & & \\
$CP_1$    &$\surd$&$\surd$&$\surd$ ($\times$2)  &$\surd$
&$\surd$& & &\\
$CP_2$    &$\surd$&$\surd$&$\surd$ & & & & $\surd$&\\
$CP_3$    &$\surd$&$\surd$&$\surd$ & & & & &$\surd$\\
$T$    & $\surd$& $\surd$& $\surd$& & & $\surd$& &\\
\end{tabular}
\end{ruledtabular}
\end{center}
\caption{\label{tab:symmetries} Quantum numbers of the eigenstates collapsing towards the same degenerate GS for each of
the ordered phases considered in this paper. When applicable, we used the standard 
notations for the  irreducible
representations of the $C_{4v}$ and $C_{2v}$ point groups, whose elements are defined
w.r.t. a plaquette center. Definitions of the $\Gamma$, $M$ and $K$ points in the Brilloin zone are given in 
Fig.~\ref{all}. ($\times 2$) denotes an additional first excited level (denoted by * in the text) in the
($M,A_1$) sector. The states with momenta $Q_B=(\pm 2\pi/3,0)$,  $(0,\pm 2\pi/3)$, $(\pi,\pm 2\pi/3)$, $(\pi,\pm 2\pi/3)$,
$Q_2=(\pm\pi/2,\pi)$,  $(\pi,\pm\pi/2)$, $Q_3=(\pm\pi/2,\pm\pi/2)$ are even under reflection w.r.t. the momentum directions.
The degeneracy of the pure columnar or plaquette (mixed) phases is 4 (8) and it is 12 for the trimerized  phase.}
\end{table}

Before showing the results of an extensive numerical analysis, we first start with a simple 
variational analysis. 
Indeed, variational ansatze for the VBC phases of Fig.~\ref{fig01}
can be easily constructed as tensor products of resonating plaquette
states (see appendix \ref{VariationalAnalysis}) and the knowledge of their relative stability provides a useful guide
for the numerical search of VBC (but is also subject to some artifact of the variational method).
For convenience, let us map the two-dimensional parameter space 
on a sphere by expressing the Hamiltonian parameters in terms of two
Euler angles $\theta$ and $\phi$, as
$ v  = \cos(\phi) \sin(\theta)$, $t_4 = \cos(\phi) \cos(\theta)$ and $t_6 = \sin(\phi)$.
The  variational phase diagram (Appendix \ref{VariationalAnalysis}) in the $(\theta,\phi)$-plane 
contains three phases;
(i) a $RK$ region,  (ii) the well-known completely isotropic 4-site plaquette
phase and (iii) a large domain covered by the {\it trimerized} VBC with a 6-site
unit cell interpolating between the columnar phase and the pure 6-site
plaquette phase. For zero loop-6
kinetic term ($\phi=0$), the phase diagram is the same than the one obtained in
\cite{trousselet} (for $\theta > -0.25 \pi$). 
Once $\phi$ is turned on, kinetic fluctuations due to 6-loop kinetic
terms suppress standard 4-site plaquette and RK phases in the vicinity of the RK point.
It is remarkable that both the $RK$ and the plaquette phase
are rather robust under the additional $t_6$ kinetic term.
However, the variational approach overestimate the stability of the RK phase (in fact limited to a single point with
algebraic dimer correlations) and of the trimerized phase (the corresponding wavefunction
has more 4-site flippable plaquettes than plaquette counterpart). In contrast, no mixed anisotropic VBC phase is found.
A careful numerical approach is therefore necessary.

\begin{figure}
\includegraphics[width=0.47\textwidth,clip]{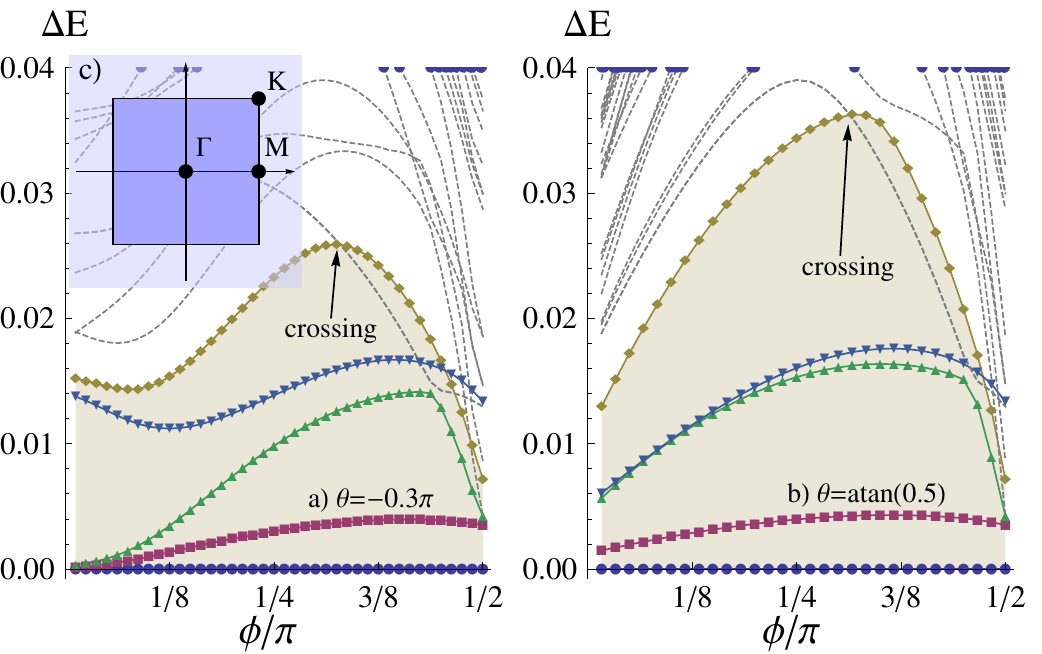}
\caption{\label{all} (Color online).  Typical ED low-energy spectra on a
$8\times 8$ cluster (the GS energy is set to zero) for a) $\theta =-0.3 \pi$
and b) $\theta =
\tan^{-1}{(0.5)}$ as a function of $\phi$.
Levels of special symmetries (see text) are highlighted as colored symbols, from bottom to top: $(\Gamma,A_1)$ (corresponding to the GS), $(M,A_1)$, $(\Gamma,B_1)$, $(K,A_1)$ and $(M,A_1)^*$. The arrow indicates the level crossing that marks the limit of the region where the later levels correspond to the lowest
excitations.
(c) Brillouin zone and its high symmetry points
$\Gamma = (0,0)$, $M = (\pi,0)$ and $K = (\pi,\pi)$.}
\end{figure}

{\it Numerical results:} We now move to ED  of clusters up to $8\times 8$ sites.
More especially, we compute the lowest energy spectrum in each symmetry
sectors, using both translations and point group symmetries. Our
analysis  is based on the symmetry classification of the tour of states in the
$(\theta,\phi)$ plane. In other words, each symmetry breaking VBC phase is characterized by a finite degeneracy of the GS with a set of well-defined
quantum numbers (see table) separated by a gap from the continuum. On a finite size cluster, the degenerate GS is split but a close inspection of the low-energy spectrum 
can provide informations on the VBC phase (if the cluster is large enough).
For $t_6=0$ (i.e. $\phi = 0$), previous results (extrapolated to the thermodynamic limit)
show that a phase transition between
the columnar phase and a mixed columnar-plaquette phase (in fact the $CP_1$ phase of this paper) occurs around
$\theta  \sim -0.03 \pi$. 
Such a phase can obtained via an {\it in phase} spontaneous dimerization in the direction of the columns of dimers of the columnar phase or, equivalently, via
a spontaneous rotation-symmetry breaking of the pure plaquette phase.
To simplify the discussion, we describe here
two representative set of parameters that contain these two phases,  $\theta =
atan(0.5)$ ($v/t_4 = 0.5$) and $\theta = -0.3\pi$, for which we have computed,
as a function of  $\phi$, the spectrum of the effective model by full ED. 
The spectra (defined w.r.t. the respective GS energies) are displayed in Fig.\ref{all}.
Special symbols have been used to label five of the six low-energy 
levels (7 states over 8) associated to the mixed $CP_1$ columnar-plaquette phase.
We donnot consider the last one, the $(K,B_1)$ level, which is believed to be more affected by finite size effects.
The plots show wide intervals of $\phi$ where the above five levels are the true lowest eigen-energies,
hence pointing towards the mixed phase as a possible GS.
The level crossing at which the low energy spectrum becomes
not anymore compatible with such a phase is indicated by an arrow. This
level crossing can be used as a first crude estimator of the range of stability of the mixed phase.
Surprisingly enough, the spectra at $\theta> 0$ is not drastically affected by a finite value of $t_6$ ($\phi>0$),
showing that the mixed phase is rather stable w.r.t an extra loop-6 term, up to
$\phi \simeq 0.3 \pi$. This range of stability will be corroborated by our
thermodynamic limit extrapolations of the order parameters (see below).
In contrast, at $\theta = -0.3 \pi$ and small $\phi$, one can see that the very lowest levels of the spectrum (i.e. those really separated by a sizable gap from the rest)
are compatible with the columnar phase whose symmetries are given in
Table.\ref{tab:symmetries}. A narrow region of mixed columnar-plaquette phase might however exist at intermediate $\phi$ values
before the level crossing involving the $(M,A_1^*)$ state occurs.
To finish this ED analysis, let us mention that, apart from the pure columnar and the mixed $CP_1$ columnar-plaquette phases, no
region in parameter space could be found where the low-energy spectrum is compatible with the other VBC phases
described above. In particular, for large (relative) $t_6$ the spectrum becomes quite dense at low energies preventing any VBC phase identification~\cite{note_cluster}.

\begin{figure}
\includegraphics[width=0.45\textwidth,clip]{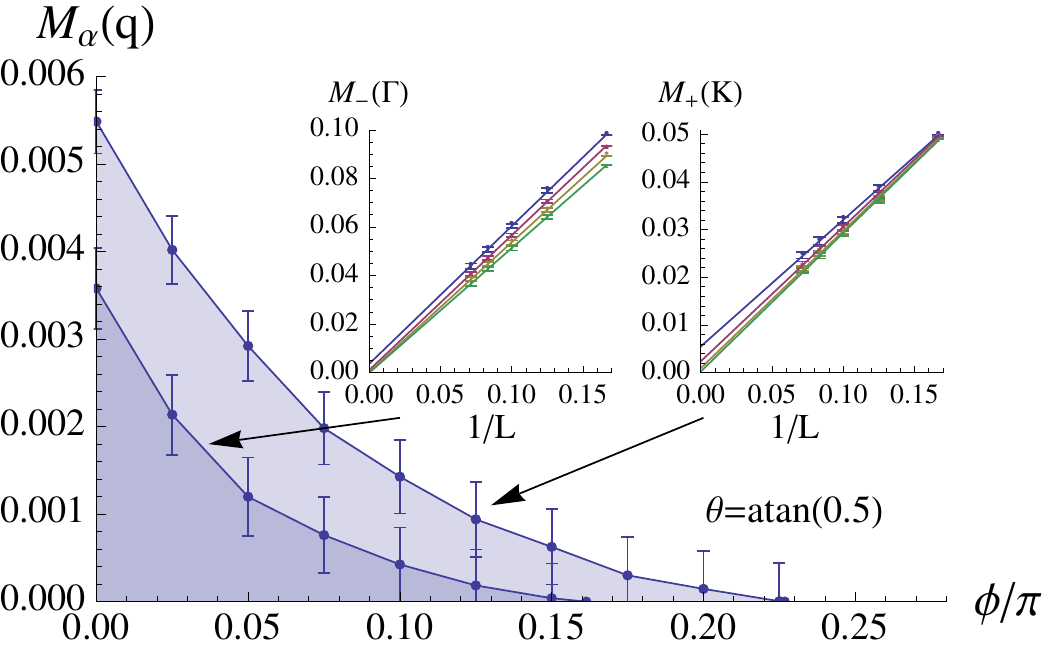}
\caption{\label{OrderParameter} (Color online). Extrapolations in the thermodynamic limit of the order
parameters defined in the text characterizing the mixed CP$_1$ columnar-plaquette phase. 
Insets:  finite
size-scaling of both $M_+(\pi,\pi)$ and $M_-(0,0)$ as a function of the inverse linear cluster size, using 36 sites ($6\times 6$), 64 sites ($8\times 8$), 100 sites ($10\times 10$), 144 sites ($12\times 12$) and 196 sites ($14\times 14$) square clusters.
The chosen value of $\theta$ corresponds to the case of the $J_1-J_2-J_3$ model studied in this work.}
\end{figure}

\begin{figure}
\includegraphics[width=0.48\textwidth,clip]{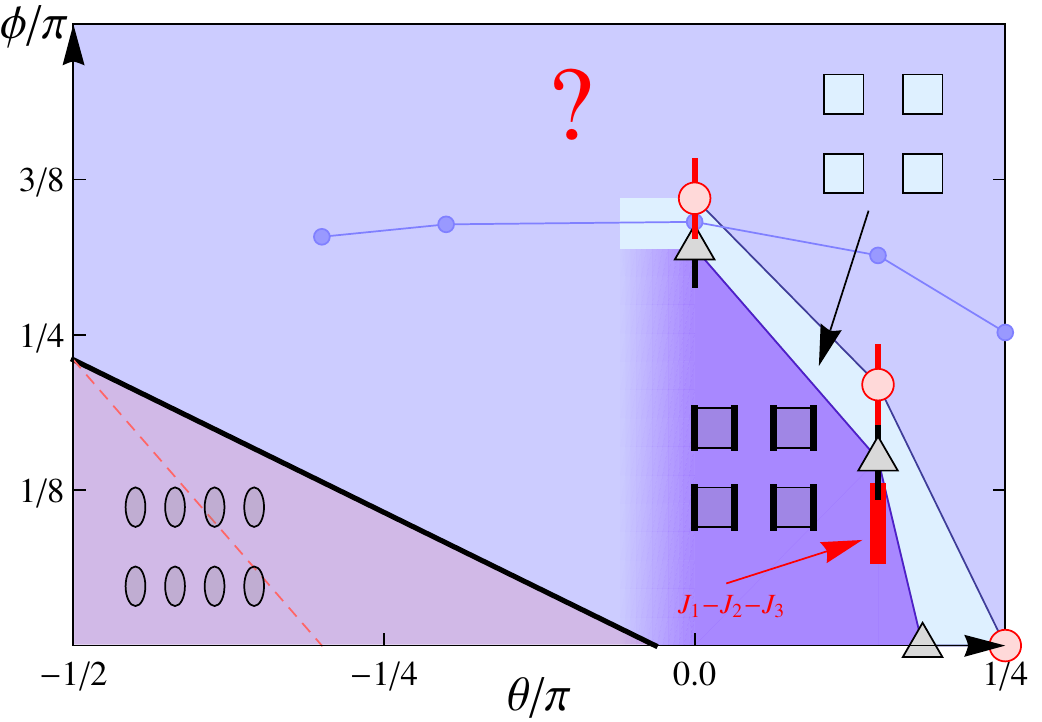}
\caption{
\label{skematicpd} 
(Color online). Phase diagram in the ($\theta$,$\phi$) plane obtained from numerical simulations. 
The previous knowledge of the $\phi=0$
case~\protect\cite{ralko1} has been used, in particular to estimate the limit of the 
columnar phase as an approximate (black) straight-line. 
The boundary of the pure-plaquette ($CP_1$ mixed) VBC phase obtained from the finite size scaling of the associated order parameter
is indicated by large circles (triangles).
The (red) thick segment corresponds to the parameter region of the frustrated quantum antiferromagnet 
with $J_2+J_3=J_1 /2$ according to the mapping described in the text.
Crude ED estimates of the boundaries of the columnar phase and of the $CP_1$ phase are indicated by a thin dashed line and by small
(blue) circles, respectively.  The (light blue) region marked by a question-mark has not been identified.}
\end{figure}

In order to give a more quantitative determination of the region of stability of the mixed $CP_1$ columnar-plaquette order, we
have computed the related plaquette structure
factors, 
\begin{eqnarray}
I_\beta(q) = \frac{\langle \Psi_0 | P_{\beta} (-q)
P_{\beta} (q) | \psi_0 \rangle}{\langle \Psi_0 | \Psi_0 \rangle},
\end{eqnarray}
 where $P_\alpha (q)$ is a {\it diagonal }
operator with the same symmetry as the degenerate GS listed in
Table.~\ref{tab:symmetries}
that we aim to target, defined as Fourier transform of plaquette
operators introduced in~[\onlinecite{ralko1}].
For positive $t_4$ and $t_6$ values, such quantities can be computed efficiently using Green Function's Quantum Monte Carlo
(GFQMC).
A Bragg-peak of $I_+(q)$ at momentum $(\pi,\pi)$ ($K$-point) 
and a divergence of $I_-(q=0)$ ($\Gamma$-point) reflect spontaneous translation and rotation symmetry 
breaking of the mixed phase, respectively.
Related order parameters $M_\beta(q) = \sqrt{I_\beta(q)}/L$ can be conveniently defined and 
results are displayed in Fig.\ref{OrderParameter} showing
size-scalings of both $M_+(\pi,\pi)$ and $M_-(0,0)$ up to cluster
size of $196$ sites. 
These data correspond to the line $v/t_4 = 0.5$, {\it i.e.} originating from the expansion of the 
microscopic model $J_1-J_2-J_3$ under  consideration here. 
Our results reveal that the Bragg peak at the $K$-point
 survives up to
$\phi < 0.3 \pi$, in good agreement with the ED criterium above. Interestingly, for increasing $\phi$,
the columnar order parameter $I_-$  vanishes before the
plaquette one: hence rotation symmetry is first recovered and a narrow region of pure plaquette order 
is stabilized between the mixed
phase region and the more complicated (unknown) phase at larger $t_6$.
The extension of the GFQMC calculation to the $\theta >0$ region which do not have ergodicity 
problems limitations, has enable to draw the phase diagram 
depicted in Fig.\ref{skematicpd}.

{\it Concluding remarks:} To finish, let us summarize our findings and their implications. 
First we have extended the QDM to the case with a finite $t_6$ amplitude for the loop-6 kinetic processes.
Although we have also extended the search for new VBC phases, we found that the previously known 
columnar, plaquette and mixed (here called CP$_1$) columnar-plaquette phases are stable when a moderate $t_6$ is added.
This conclusion is first obtained by a close inspection of the low-energy spectrum of the model on finite clusters
(ED). The quantitative determination of the extensions of the three previous phases is made possible by
GFQMC simulations which do not suffer from the sign problem when $t_6>0$. Still, it has not been possible to characterize
the GS in the whole parameter space, in particular when $t_6$ dominates (accumulation of low-energy states)
or when  $v<0$ (i.e. $\theta <0$) where the GFQMC encounters numerical
limitations (both cases being physically irrelevant anyway).
Our present work on the generalized QDM turns out to be very useful to make progress in the understanding of the frustrated S=1/2 quantum antiferromagnet on the square lattice.
Indeed, it was previously argued that, in the vicinity of $J_2\simeq J_3\simeq J_1/4$ and along the $J_2+J_3=J_1/2$ line,
 a truncation in the NN SU(2)-singlet basis was legitimate. Using this argument and generalizing the RK expansion in terms of the magnitudes of the overlaps 
 of the elements of the truncated (non-orthogonal) basis we have made a link between the microscopic model and some small region of parameter space of the
 generalized QDM where, fortunately, a precise characterization of the phase can be made. We can therefore deduce that the frustrated S=1/2 quantum antiferromagnet
 exhibits in the vicinity of $J_2\simeq J_3\simeq J_1/4$ the same type of lattice symmetry breaking as the mixed columnar-plaquette phase ($CP_1$) of the QDM.
 In the language of the quantum antiferromagnet, it is a eight-fold degenerate SU(2)-symmetric phase with a $2\times 2$ unit cell (like the plaquette phase)
 and rotation symmetry breaking (like the columnar phase). While a previous investigation of the microscopic model on a $6\times 6$ cluster 
 has indeed provided evidence of plaquette correlations \cite{mambrini}, only the mapping to the effective model (which can be studied on much larger clusters) provides enough accuracy 
 to show the spatially anisotropic nature of this spin-singlet plaquette VBC phase.

\begin{acknowledgements}
 D.P. and A.R. are indebted to F. Becca for useful discussions at an early stage of this work.
\end{acknowledgements}

\appendix
\section{Variational Analysis} \label{VariationalAnalysis}
\begin{figure}[h]
\includegraphics[width=0.47\textwidth,clip]{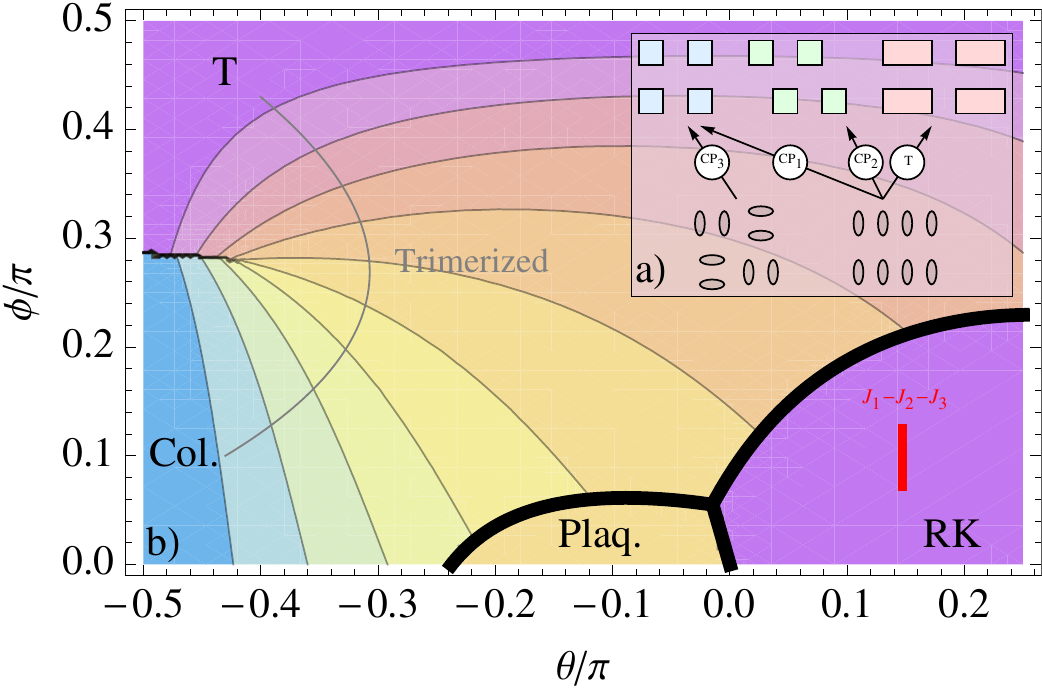}
\caption{\label{fig05}  (a) Variational wave functions (associated to their respective VBC states) considered in this Appendix.  
Generalized anisotropic VWF labeled by $|A_a\big>$ and $|B\big>$ 
interpolate between the most symmetric limits shown on the figure.
(b) Variational phase diagram as a function of $\theta$ (x-axis) and $\phi$
(y-axis).
The thick red line shows the region of parameters of the effective model 
that maps to the $J_1-J_2-J_3$ model along the $J_2+J_3=J_1/2$ line for which
the ground state is well described by
singlet coverings.}
\end{figure}

The effective generalized Quantum Dimer Model on the square lattice 
originated from the microscopic frustrated Heisenberg Hamiltonian studied here can be first investigated by a 
variational method. 
Indeed, it is possible
to construct variational ansatze for Valence Bond Crystal phases (see inset of Fig.~\ref{fig05})
as simple tensor products of resonating plaquette states,
extending the number of possible phases arising in the standard QDM. Variational energies can then be computed analytically.
Despite its simplicity, this
approach reveals itself to be a useful guide for the numerical search of VBC
(although artifacts due to its variational nature are expected) and 
can easily incorporate the symmetry analysis provided in the paper.
In other words, there is a one-to-one correspondence between these variational wave functions (VWF)
and the VBC states defined in the paper by the set of spontaneously broken symmetries.

We shall consider a set of five different VWF, (i) the well known
$|RK\rangle$ one provided by Rokhsar and Kivelson~\cite{rokhsar} as an equal weight superposition of all 
dimer configurations, (ii) three VWF based on 4-site plaquette tensor products
($| CP_i
\rangle$, $i=1,2,3$) and (iii) one with
a 6-site unit cell interpolating between the columnar phase and the pure
6-site plaquette phase ({\it e.g.} $|T\rangle$ in Fig.~\ref{fig05}). Excepted
for $| RK \rangle$, all these VWF depend on a parameter $\alpha$ which allows
a continuous interpolation between pure singlet crystals and highly resonating
VBC. To illustrate this construction, here is the expression of one of the anisotropic 
4-site plaquette phases and the above mentioned 6-site plaquette one:
\begin{eqnarray}
 | CP_1 \rangle &=& \prod_{\otimes p} \cos(\alpha)
 |\includegraphics[width=0.018\textwidth,clip]{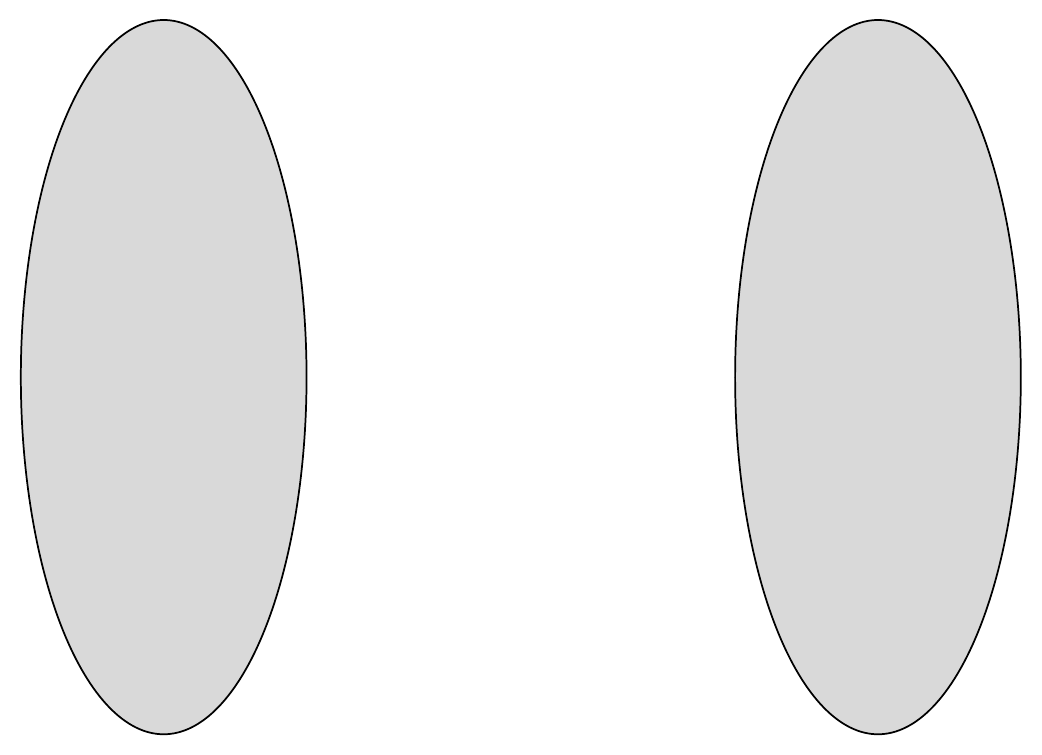}\rangle_p +\sin(\alpha)
 |\includegraphics[width=0.012\textwidth,clip]{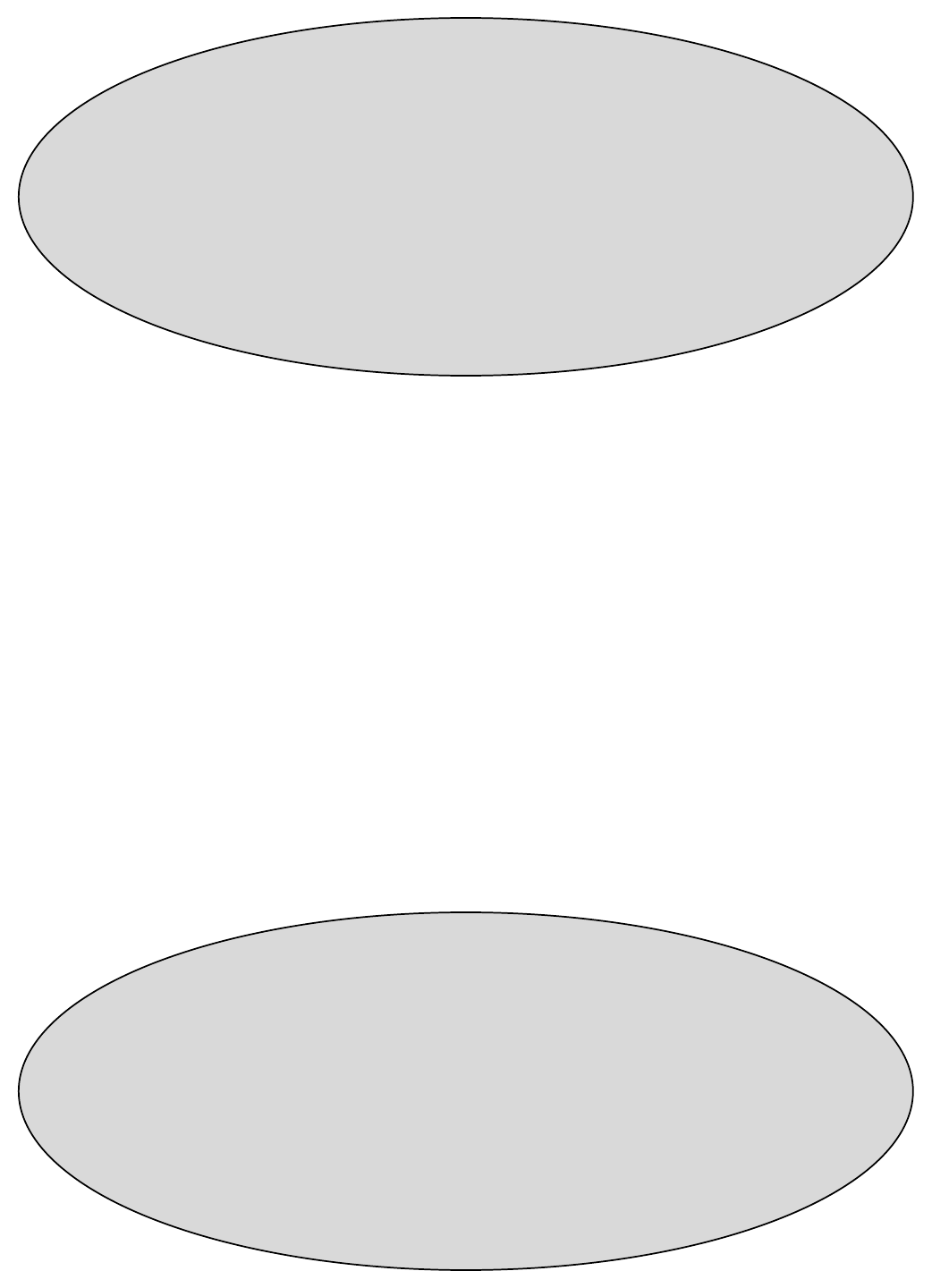}\rangle_p \\
| T \rangle &=& \prod_{\otimes p'}\sin(\alpha)
|\includegraphics[width=0.03\textwidth,clip]{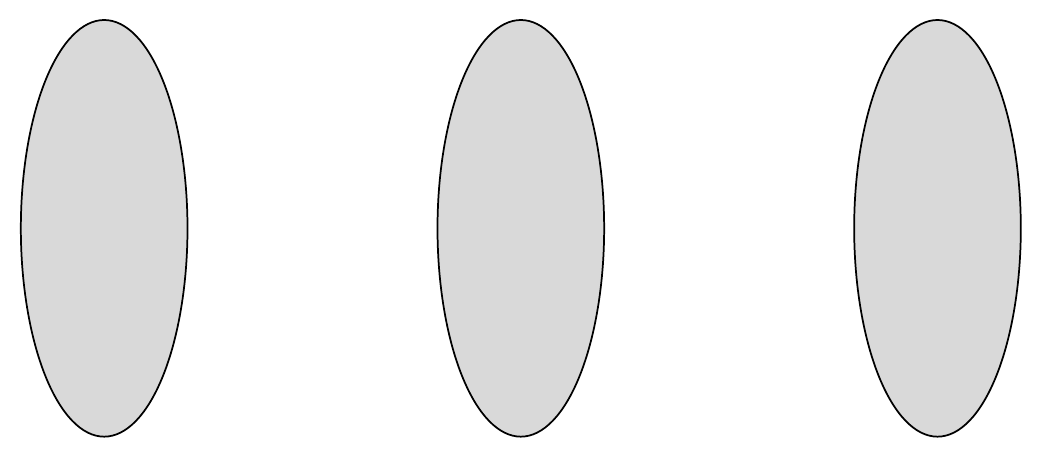}\rangle_{p'}
+\frac{\cos(\alpha)}{\sqrt{2}} \left(
|\includegraphics[width=0.027\textwidth,clip]{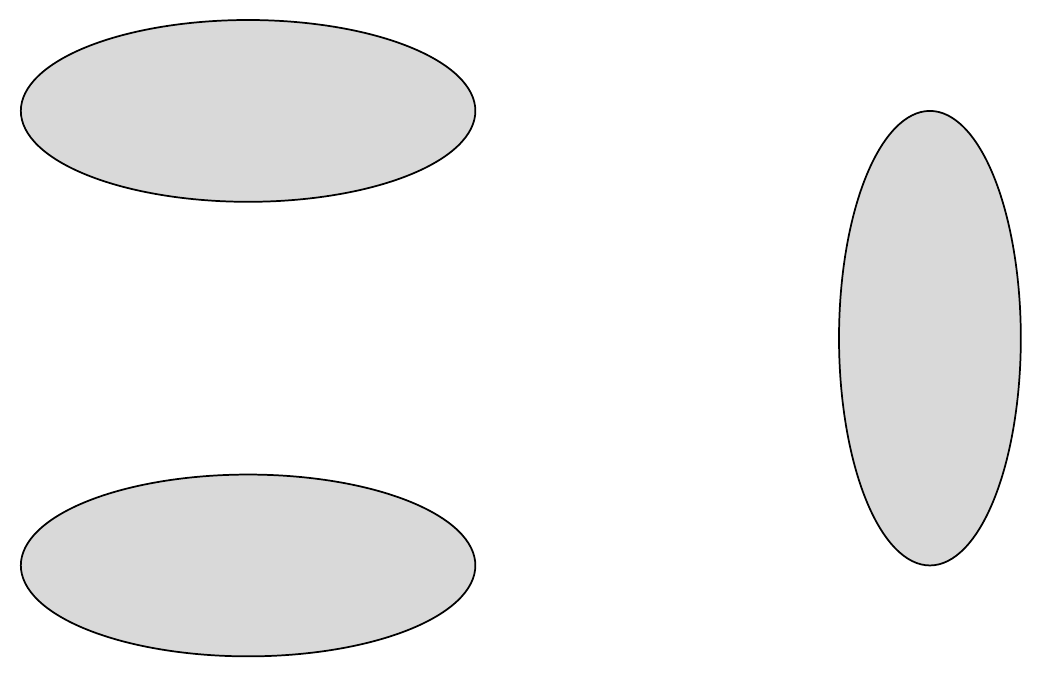} \rangle_{p'} +
|\includegraphics[width=0.027\textwidth,clip]{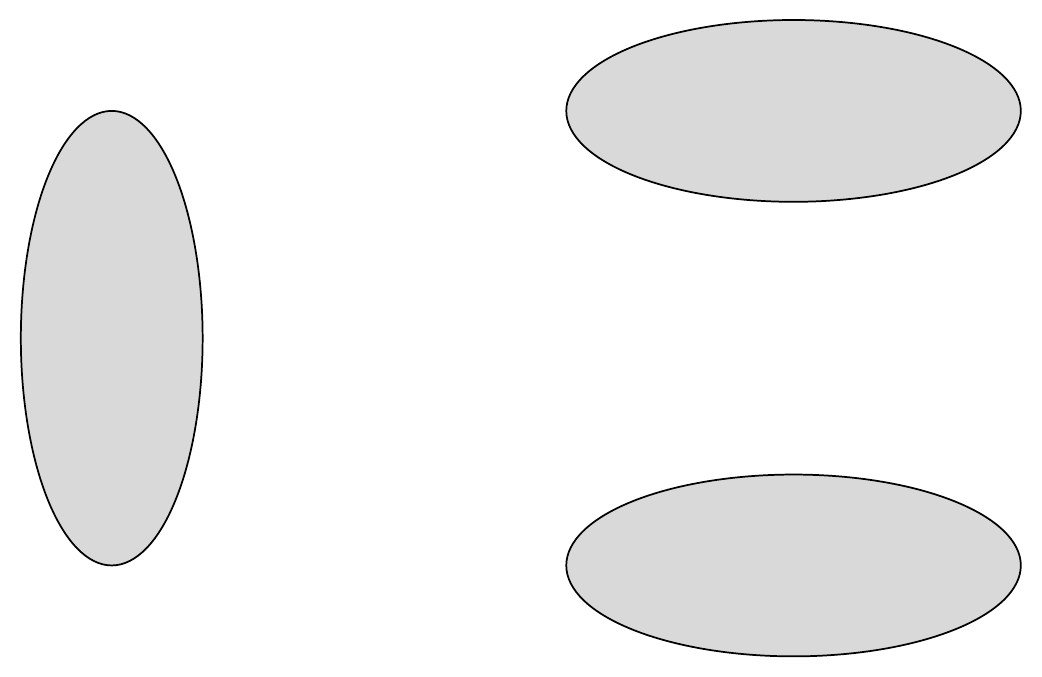} \rangle_{p'} \right), \nonumber
\end{eqnarray}
where the product is performed over the set of separate plaquettes $p$ or $p'$ as suggested in Fig.~\ref{fig05}.

In order to describe the stability of these phases w.r.t. the parameters, the
expectation value of the effective generalized QDM Hamiltonian is computed. The only
off-diagonal terms of
${\cal H}$ contributing to $\langle \Gamma | {\cal H} | \Gamma \rangle $, with
$| \Gamma \rangle$ being one of those VWF, are plaquette flips on occupied plaquettes
and the diagonal potential term. This leads to contributions proportional to
$\cos(\alpha)\sin(\alpha)$ for all the VWF, plus one in $\cos^2(\alpha)$ for
the 6-site plaquette one. For the diagonal terms, both occupied and
non-occupied plaquettes yield non-zero contributions to the expectation value
of ${\cal H}$. It is worth to emphasize that the $| RK \rangle$ WF requires a
special analysis. Using well-defined Pfaffian techniques, one can compute
analytically the probability of flipping a 4-site and a 6-site plaquette,
which are respectively $P_4 = 1/4$ and $P_6 = 0.0330112(1)$.
Finally, these expectation values, as a function of the Hamiltonian parameters
$v(\theta,\phi)$, $t_4(\theta,\phi)$ and $t_6(\theta,\phi)$, are given by:
\begin{eqnarray}
 E_{RK} &=& v - t_4 - 4 t_6 P_6 \nonumber \\
 E_{CP_1} &=& v(1+\cos^4(\alpha)+\sin^4(\alpha)) - 2 t_4
\cos(\alpha)\sin(\alpha)\nonumber \\
 E_{CP_2} &=& v(1+\cos^4(\alpha)) -2 t_4
\cos(\alpha)\sin(\alpha)\nonumber \\
 E_{CP_3} &=& v(1+2 \cos^2(\alpha)\sin^2(\alpha)) -2 t_4
\cos(\alpha)\sin(\alpha)\nonumber \\
 E_{T} &=& v(2\cos^2(\alpha)+4\sin^2(\alpha) \nonumber \\ &&
+3\cos^4(\alpha)/2+2\sin^4(\alpha))/3 \nonumber \\ &-& t_4
\frac{8\cos(\alpha)\sin(\alpha)}{3\sqrt{2}}  
- t_6 \frac{2\cos^2(\alpha)}{3}.
\end{eqnarray}
These energies are then minimized w.r.t. $\alpha$ and the variational phase
diagram displayed in Fig.~\ref{fig05} is obtained in the $(\theta,\phi)$-plane.
This phase diagram is discussed in the paper. The colors correspond to
different values of the $\alpha$ parameter, $\alpha = \pi/2$ (blue) for the
pure columnar phase, $\alpha = 0$ (purple) for the pure 6-site resonating
plaquette phase and $\alpha = \pi/4$ (orange) for the isotropic 4-site plaquette
phase. The RK region has an arbitrary color.

\begin{figure}
\includegraphics[width=0.48\textwidth,clip]{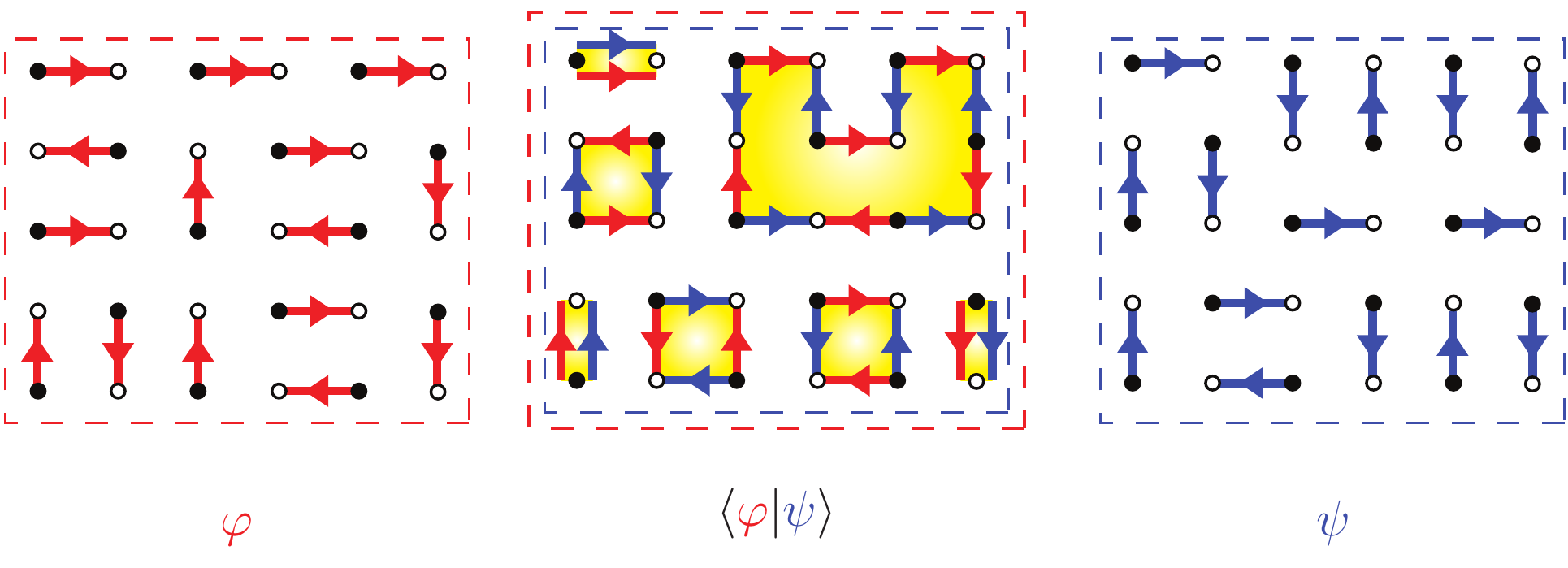}
\caption{
\label{fig:overlaps} 
(Color online). Overlap $\langle \varphi | \psi \rangle$ between two VB configurations $\varphi$ ad $\psi$. Closed loops that appear in the overlap digram $g$ are represented as yellow shades.}
\end{figure}
\section{Derivation of the Hamiltonian} \label{Derivation}

This appendix is devoted to a technical description of the generalized QDM derivation scheme. 

{\it Choice of a small parameter \& disconnected processes.} As briefly described in the article, the generalized QDM is obtained by developping ${\cal H}^{\textrm{eff}}={\cal O}^{-1/2} {\cal H} {\cal O}^{-1/2}$ according to the matrix element hierarchy of both ${\cal O}$ and ${\cal H}$ in the VB basis. Indeed, the amplitude of ${\cal O}_{\varphi,\psi}$ (with $\varphi$ and $\psi$ two VB configurations) only depends on the {\em number of loops} $n_l$ of the overlap diagram $g=(\varphi,\psi)$ obtained by superimposing the two configurations : ${\cal O}_g= \alpha^{N - 2n_l(g)}$ with $\alpha=1/\sqrt{2}$ and $N$ the total number of sites (see figure \ref{fig:overlaps}). 

The maximal number of loops is obtained for $\varphi=\psi$. In this case $g$ is just a collection of $N/2$ trivial loops with lentgh ${\cal L}_l=2$ and ${\cal O}_g=1$. The next term is obtained by considering one non-trivial loop with length ${\cal L}_l=4$ (\diag[g_1_1]{0.045}) while all remaining $(N-4)/2$ loops are chosen trivial which leads to ${\cal O}_g=\alpha^2$. In the same spirit, one  ${\cal L}_l=6$ loop (\diag[g_2_2]{0.06}) and $(N-6)/2$ trivial loops is an $\alpha^4$-order process (see the fourth line of tables \ref{tab:expansion} and \ref{tab:native_terms}). Such a construction suggests that a quite natural driving parameter for the expansion is the length ${\cal L}_l$ of a unique non-trivial loop surrounded by $(N-{\cal L}_l)/2$ trivial loops : such a process indeed appears at order $\alpha^{{\cal L}_l - 2}$. However the total length of loops is a constraint quantity ($\sum_{l \in g} {\cal L}_l = N$) and other (non connected) terms indeed appear in $\cal O$ at the same order. For example, \diag[g_2_1]{0.075} formed by two disconnected squares also occurs in ${\cal O}$ with the amplitude $\alpha^4$ despite the fact the non-trivial contour length is different from e.g. \diag[g_2_2]{0.06}.

For this reason, in the derivation scheme presented here, we chose the {\em overall number of loops} as the driving parameter for the expansion of ${\cal O}$ rather than the commonly used {\em length of the loops}~\cite{rokhsar,ralko2,moessner}. The key difference lies in the presence of disconnected processes such as \diag[g_2_1]{0.075} : as we will see, they cancel at every order of the final effective hamiltonian, but are crucial in the calculation because they are responsible for the emergence of non trivial diagonal and off-diagonal connected processes.

In the expansion of ${\cal O}$, we use the following notation,
\begin{equation}
	\label{eq:overlap_expansion}
	{\cal O} = \sum_{p \geq 0} \alpha^{2p} \omega_p,
\end{equation}
where $\omega_p$ are combinations of $\omega_p^g$ process on graphs $g$ :
\begin{equation}
	\label{eq:overlap_graphs}
	\omega_p = \sum_{g} \omega_p^g.
\end{equation}

For a full list of $\omega_p^g$ up to order $\alpha^8$, see table \ref{tab:native_terms}.

\begin{table}[htbp]
\begin{center}
\begin{tabular}{|c||c|c|}\hline\hline
  Processes &  $\cal O$ & ${\cal H}$ \\ \hline\hline
 \Addsp[4mm]{0.} Id & $1$ & $0$ \\ \hline\hline
 \Addsp[6mm]{0.} \diag[g_1_1]{0.06}  & $\alpha ^2$ & $2 \left(J_2-J_1\right) \alpha ^2$ \\ \hline\hline
 \Addsp[6mm]{0.} \diag[g_2_1]{0.10} & $\alpha ^4$ & $4 \left(J_2-J_1\right) \alpha ^4$ \\ \hline
 \Addsp[6mm]{0.} \diag[g_2_2]{0.08} & $\alpha ^4$ &  $2 \left(-2 J_1+2 J_2+J_3\right) \alpha ^4$  \\ \hline\hline
 \Addsp[6mm]{0.} \diag[g_3_1]{0.14} & $\alpha ^6$ & $6 \left(J_2-J_1\right) \alpha ^6$ \\ \hline
  \Addsp[6mm]{0.} \diag[g_3_2]{0.12} & $\alpha ^6$ &  $2 \left(-3 J_1+3 J_2+J_3\right) \alpha ^6$ \\ \hline
\Addsp[8mm]{0.} \diag[g_3_3]{0.08}  & $\alpha ^6$ & $2 \left(-3 J_1+3 J_2+2 J_3\right) \alpha ^6$ \\ \hline
 \Addsp[6mm]{0.} \diag[g_3_4]{0.10} & $\alpha ^6$ & $2 \left(-3 J_1+3 J_2+2 J_3\right) \alpha ^6$ \\ \hline\hline
 \Addsp[6mm]{0.} \diag[g_4_1]{0.18} & $\alpha ^8$ & $8 \left(J_2-J_1\right) \alpha ^8$ \\ \hline
 \Addsp[6mm]{0.} \diag[g_4_2]{0.16} & $\alpha ^8$ & $2 \left(-4 J_1+4 J_2+J_3\right) \alpha ^8$ \\ \hline
 \Addsp[6mm]{0.} \diag[g_4_3]{0.14} & $\alpha ^8$ & $4 \left(-2 J_1+2 J_2+J_3\right) \alpha ^8$ \\ \hline
 \Addsp[6mm]{0.} \diag[g_4_4]{0.14} & $\alpha ^8$ & $4 \left(-2 J_1+2 J_2+J_3\right) \alpha ^8$ \\ \hline
 \Addsp[8mm]{0.} \diag[g_4_5]{0.12} & $\alpha ^8$ & $4 \left(-2 J_1+2 J_2+J_3\right) \alpha ^8$ \\ \hline
\Addsp[6mm]{0.} \diag[g_4_6]{0.12} & $\alpha ^8$ & $2 \left(-4 J_1+4 J_2+3 J_3\right) \alpha ^8$ \\ \hline
\Addsp[8mm]{0.} \diag[g_4_7]{0.10} & $\alpha ^8$ & $2 \left(-4 J_1+4 J_2+3 J_3\right) \alpha ^8$ \\ \hline
\Addsp[8mm]{0.} \diag[g_4_8]{0.10} & $\alpha ^8$ & $2 \left(-4 J_1+4 J_2+3 J_3\right) \alpha ^8$ \\ \hline
\Addsp[8mm]{0.} \diag[g_4_9]{0.10} & $\alpha ^8$ & $2 \left(-4 J_1+4 J_2+3 J_3\right) \alpha ^8$ \\ \hline\hline
\end{tabular}
\end{center}
\caption{\label{tab:native_terms} Expansions of $\cal O$ and $\cal H$ up to order $\alpha^8$.}
\end{table}

\begin{figure}
\includegraphics[width=0.3\textwidth,clip]{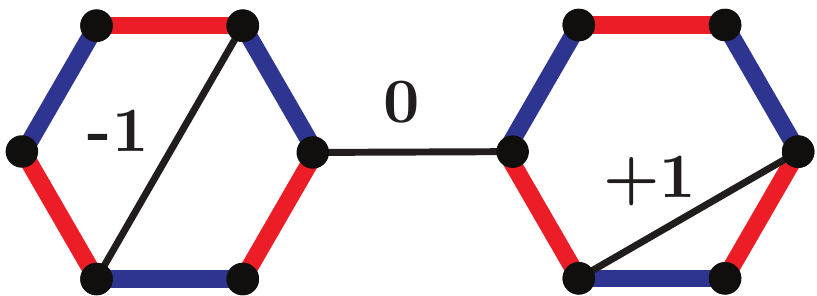}
\caption{
\label{fig:HamiltonianTerms} 
(Color online).  The matrix element $\langle \varphi | (4/3){\bf S}_i . {\bf S}_j | \psi \rangle$ expressed as $f_{ij} \langle \varphi |\psi \rangle$. 
Bond $(i,j)$ is represented as a solid black line. Red and blue bonds represent $| \varphi \rangle $ and $| \psi \rangle$. $f_{ij}= +1$ (respectively $f_{ij}=-1$) if $i$ and $j$ belong to the same loop at even (respectively odd) distance along the loop. $f_{ij}=0$ if $(i,j)$ connects two sites on distinct loops (see Ref.~\protect\onlinecite{Sutherland}). 
 }
\end{figure}

\begin{table*}[htbp]
\begin{center}
\begin{tabular}{|c||c|c|c|c|} \hline\hline
 \multirow{3}{*}{Processes}				  &	\multicolumn{4}{|c|}{${\cal H}^{\textrm{eff}}={\cal O}^{-1/2} {\cal H} {\cal O}^{-1/2}$} \\ \cline{2-5}
 																		& Analytic  & $J_2=J_1/2$ 	& $J_2=J_1/4$ & $J_3=J_1/2$				\\ 
 & expression & $J_3=0$ & $J_3=J_1/4$ & $J_2=0$																		\\ \hline\hline
 \Addsp[9mm]{0.} \diag[g_1_2]{0.045} & $2 \left(J_1-J_2\right) \alpha ^4 \left(1+\alpha ^4\right)$	& 0.625 & 0.46875 & 0.3125 \\ \hline
 \Addsp[9mm]{0.} \diag[g_1_1]{0.045} & $-2 \left(J_1-J_2\right) \alpha ^2 \left(1+\alpha ^4\right)$	& -1.25 & -0.9375 & -0.625 \\ \hline
 \Addsp[9mm]{0.} \diag[g_2_2]{0.06} & $-2  \left(\left(J_1-J_2\right) \left(1+\frac{5}{4} \alpha ^4\right)-J_3 \left(1+\frac{1}{4}\alpha ^4\right)\right)\alpha ^4$	& -0.320312 & -0.304687 & -0.289062 \\ \hline
 \Addsp[12mm]{0.} \diag[g_3_3]{0.06} & $2 \left(-J_1+J_2+J_3\right) \alpha ^6$	& -0.125 & -0.125 & -0.125 \\ \hline
 \Addsp[9mm]{0.} \diag[g_3_4]{0.075} & $\left(-J_1+J_2+2 J_3\right) \alpha ^6$	& 0. & -0.03125 & -0.0625 \\ \hline
 \Addsp[9mm]{0.} \diag[g_2_4]{0.06} & $\left(J_1-J_2-J_3\right) \alpha ^6$	& 0.0625 & 0.0625 & 0.0625 \\ \hline
 \Addsp[9mm]{0.} \diag[g_2_6]{0.06} & $\left(3 J_1-3 J_2-J_3\right) \alpha ^8$	& 0.15625 & 0.125 & 0.09375 \\ \hline
 \Addsp[9mm]{0.} \diag[g_2_7]{0.06} & $\frac{1}{2} \left(-5 J_1+5 J_2+J_3\right) \alpha ^8$	& -0.140625 & -0.109375 & -0.078125 \\ \hline
 \Addsp[9mm]{0.} \diag[g_2_8]{0.06} & $4 \left(J_1-J_2\right) \alpha ^8$	& 0.25 & 0.1875 & 0.125 \\ \hline
 \Addsp[12mm]{0.} \diag[g_3_8]{0.06} & $\frac{3}{4} \left(J_1-J_2-J_3\right) \alpha ^8$	& 0.0234375 & 0.0234375 & 0.0234375 \\ \hline
 \Addsp[9mm]{0.} \diag[g_3_9]{0.075} & $\frac{1}{4} \left(2 J_1-2 J_2-3 J_3\right) \alpha ^8$	& 0.0078125 & 0.0117187 & 0.015625 \\ \hline
 \Addsp[9mm]{0.} \diag[g_3_11]{0.075} & $\frac{1}{4} \left(6 J_1-6 J_2-5 J_3\right) \alpha ^8$	& 0.0546875 & 0.0507812 & 0.046875 \\ \hline
 \Addsp[9mm]{0.} \diag[g_3_12]{0.075} & $\frac{1}{4} \left(7 J_1-7 J_2-3 J_3\right) \alpha ^8$	& 0.0859375 & 0.0703125 & 0.0546875 \\ \hline
 \Addsp[9mm]{0.} \diag[g_4_6]{0.09} & $\left(3 J_1-3 J_2+J_3\right) \alpha ^8$	& 0.21875 & 0.15625 & 0.09375 \\ \hline
 \Addsp[12mm]{0.} \diag[g_4_7]{0.075} & $-\frac{3}{4} \left(J_1-J_2-2 J_3\right) \alpha ^8$	& 0. & -0.0117187 & -0.0234375 \\ \hline
 \Addsp[12mm]{0.} \diag[g_4_8]{0.075} & $\frac{1}{4} \left(-5 J_1+5 J_2+6 J_3\right) \alpha ^8$	& -0.03125 & -0.0351562 & -0.0390625 \\ \hline
 \Addsp[12mm]{0.} \diag[g_4_9]{0.075} & $\frac{1}{2} \left(3 J_1-3 J_2+4 J_3\right) \alpha ^8$	& 0.15625 & 0.101562 & 0.046875 \\ \hline\hline
\end{tabular}
\end{center}
\caption{\label{tab:expansion2} Expansion of ${\cal
H}^{\textrm{eff}}$ up to order $\alpha^8$.}
\end{table*}

{\it Heisenberg hamiltonian expansion.} The action of each term of the Heisenberg hamiltonian (\ref{eq:heisenberg}) of a VB state consist in a reconfiguration of dimers. Thus, $\langle \varphi | {\cal H} | \psi \rangle$ can be deduced form inspecting the topology of the overlap diagram $\langle \varphi | \psi \rangle$
as recalled in figure \ref{fig:HamiltonianTerms}. This allow ${\cal H}$ to be expanded in power of $\alpha$ (see Ref.~\onlinecite{Sutherland}) similarly to $\cal O$ : 

\begin{equation}
	\label{eq:hamiltonian_expansion}
	{\cal H} = \sum_{p \geq 0}  \alpha^{2p} \sum_{g} h_p^g \omega_p^g.
\end{equation}

Note that it is convenient here to rescale the hamiltonian by a factor $4/3$. Furthermore, evaluating  $\langle \varphi | (4/3){\bf S}_i . {\bf S}_j | \psi \rangle$ generically involves an extensive number of lentgh-2 loops which only effect is to produce a trivial extensive contribution to the matrix element. This can be removed by scaling and shifting ${\cal H}$ to $(4/3){\cal H}+J_1 N/2$. Using this convention, the expansion of ${\cal H}$ contains only kinetic terms. For a full list of $h_p^g$ up to order $\alpha^8$, see the last column of table \ref{tab:native_terms}.

{\it Fusion rules \& effective hamiltonian.} The first step to compute the effective hamiltonian ${\cal H}^{\textrm{eff}}={\cal O}^{-1/2} {\cal H} {\cal O}^{-1/2}$ is to derive the expression of ${\cal O}^{-1/2}$. To achieve this, we use the formal expansion :
\begin{equation}
\label{eq:formal_expansion}
{\cal O}^{\tau} = \sum_{k \geq 0} \frac{\Gamma(1+\tau)}{\Gamma(1+\tau-k)\Gamma(1+k)} \left ( {\cal O} - 1\right )^k
\end{equation}
Powers of ${\cal O}$ and products with ${\cal H}$ generically involve symmetric products of diagrams (see table \ref{tab:native_terms}) that do not commute, e.g $\left \{ \diag[g_1_1]{0.045}, \diag[g_2_2]{0.06} \right \} = \diag[g_1_1]{0.045} \otimes \diag[g_2_2]{0.06}+ \diag[g_2_2]{0.06}\otimes \diag[g_1_1]{0.045}$. Evalutating these products requires establishing {\em fusion rules} that (i) governs algebraic properties of diagrams and (ii) generate, order by order, new diagonal and off-diagonal processes. We give below, the minimal set of rule that are needed up to order $\alpha^8$.

{\it Order 4 fusion rule :}
	\begin{equation*}
		\frac{1}{2} \left \{ \diag[g_1_1]{0.045},\diag[g_1_1]{0.045} \right \} = \diag[g_1_2]{0.045}+\diag[g_2_2]{0.06}+2\diag[g_2_1]{0.075}
	\end{equation*}
	
{\it Order 6 fusion rules :}
	\begin{align*}
		\left \{ \diag[g_1_1]{0.045},\diag[g_2_1]{0.075}\right \} &= 2\;\diag[g_2_3]{0.075}+6\;\diag[g_3_1]{0.105}+2\;\diag[g_3_2]{0.09}+\diag[g_3_4]{0.075}\\
		\left \{ \diag[g_1_1]{0.045},\diag[g_1_2]{0.045}\right \} &= 2\;\diag[g_1_1]{0.045}+2\;\diag[g_2_3]{0.075}+\diag[g_2_4]{0.06} \\
		\left \{ \diag[g_1_1]{0.045},\diag[g_2_2]{0.06}\right \} &= \diag[g_2_4]{0.06}+2\;\diag[g_3_2]{0.09}+2\;\diag[g_3_3]{0.06}+2\;\diag[g_3_4]{0.075}
	\end{align*}

{\it Order 8 fusion rules :}
	\begin{align*}
		\frac{1}{2} \left \{ \diag[g_1_2]{0.045},\diag[g_1_2]{0.045} \right \}	&= \diag[g_1_2]{0.045} + 2\;\diag[g_2_5]{0.075} + 2\;\diag[g_2_8]{0.06} \\
		\left \{ \diag[g_1_2]{0.045},\diag[g_2_2]{0.06} \right \}  						&= 2\;\diag[g_3_7]{0.09} +\diag[g_3_8]{0.06} +\diag[g_3_9]{0.075} \\
		\left \{ \diag[g_1_2]{0.045} , \diag[g_2_1]{0.075} \right \} 						&= 4\;\diag[g_2_1]{0.075} + 2\;\diag[g_3_5]{0.105} +\diag[g_3_6]{0.09} + \diag[g_3_11]{0.075} \\
		\frac{1}{2} \left \{ \diag[g_2_1]{0.075},\diag[g_2_1]{0.075} \right \}  &= \diag[g_2_5]{0.075}+2\;\diag[g_3_5]{0.105}+\diag[g_3_7]{0.09}+6\;\diag[g_4_1]{0.135} \\
																																					&+2\;\diag[g_4_2]{0.12}+\diag[g_4_3]{0.105}+\diag[g_4_4]{0.105}+\diag[g_4_6]{0.09} \nonumber \\
		\frac{1}{2} \left \{ \diag[g_2_2]{0.06},\diag[g_2_2]{0.06} \right \} 	&=\diag[g_2_6]{0.06}+\diag[g_3_12]{0.075}+2\;\diag[g_4_3]{0.105}+\diag[g_4_6]{0.09} \\ &+\diag[g_4_7]{0.075}+\diag[g_4_9]{0.075} \\
		\left \{ \diag[g_2_1]{0.075},\diag[g_2_2]{0.06} \right \} 							&=\diag[g_3_6]{0.09}+\diag[g_3_9]{0.075}+2\;\diag[g_4_2]{0.12}+2\;\diag[g_4_4]{0.105} \\
																																&+2\;\diag[g_4_5]{0.09}+\diag[g_4_6]{0.09}+\diag[g_4_7]{0.075}+\diag[g_4_8]{0.075}+\diag[g_4_9]{0.075} \nonumber \\
		\left \{ \diag[g_1_1]{0.045}, \diag[g_3_1]{0.105} \right \}							&=2\;\diag[g_3_5]{0.105}+8\;\diag[g_4_1]{0.135}+2\;\diag[g_4_2]{0.12}+\diag[g_4_4]{0.105} \\
		\left \{ \diag[g_1_1]{0.045},\diag[g_3_2]{0.09} \right \}	 						&=\diag[g_3_6]{0.09}+2\;\diag[g_3_7]{0.09}+4\;\diag[g_4_2]{0.12}+4\;\diag[g_4_3]{0.105} \\
																																					&+2\;\diag[g_4_4]{0.105}+2\;\diag[g_4_5]{0.09}+2\;\diag[g_4_6]{0.09}+\diag[g_4_7]{0.075} \nonumber \\
		\left \{ \diag[g_1_1]{0.045},\diag[g_3_4]{0.075} \right \}							&=\diag[g_3_9]{0.075}+\diag[g_3_11]{0.075}+2\;\diag[g_4_4]{0.105}\\ &+2\;\diag[g_4_6]{0.09}+\diag[g_4_7]{0.075}+\diag[g_4_8]{0.075}  \\
		\left \{\diag[g_1_1]{0.045},\diag[g_3_3]{0.06} \right \} 							&=\diag[g_3_8]{0.06}+2\;\diag[g_4_5]{0.09}+\diag[g_4_7]{0.075}+2\;\diag[g_4_8]{0.075}+2\;\diag[g_4_9]{0.075} \\
  	\left \{\diag[g_1_1]{0.045},\diag[g_2_3]{0.075} \right \} 						&=4\;\diag[g_2_1]{0.075}+4\;\diag[g_2_5]{0.075}+4\;\diag[g_3_5]{0.105}+\diag[g_3_6]{0.09}\\
  	&+2\;\diag[g_3_7]{0.09}+\diag[g_3_9]{0.075} \\
  	\left \{\diag[g_1_1]{0.045},\diag[g_2_4]{0.06} \right \} 							&=\diag[g_2_2]{0.06}+2\;\diag[g_2_7]{0.06}+2\;\diag[g_3_6]{0.09}+\diag[g_3_8]{0.06}+\diag[g_3_9]{0.075}+\diag[g_3_10]{0.075} \\
  	\diag[g_3_10]{0.075} 																									&=\diag[g_3_11]{0.075}+\diag[g_3_12]{0.075}
  	\end{align*}

Note that contractions of diagrams not only produce potential (diagonal) terms, but also non-trivial assisted off-diagonal operators, such as e.g. \diag[g_2_4]{0.06} that flip a plaquette only if a dimer is present next to it.
Special process also appear such as \diag[g_3_11]{0.075} (respectively \diag[g_3_12]{0.075}) which simultaneously flip two neighboring plaquettes with parallel (respectively perpendicular) dimers. In table \ref{tab:expansion2}, we summurize the result of the expansion up to order $\alpha^8$. Interestingly enough, all disconnected terms vanish after simplifying the product ${\cal O}^{-1/2} {\cal H} {\cal O}^{-1/2}$. The demonstration of this generic property is beyond the scope of the present paper and will be presented elsewhere~\cite{MambriniSchwandt}. At this level, let us remark that this absence of non-local terms in the generalized QDM Hamiltonian is physically satisfactory and is a strong indication of the internal consistency
of the derivation scheme presented here.




\begin{thebibliography}{99}
\bibitem{note_VB} A Valence Bond is a (spatially anti-symmetric) singlet pair of two S=1/2 spins.
\bibitem{note_cluster} Note that the trimerized phase is frustrated on the $8\times 8$ cluster. However, investigation of $6\times 6$ and $6\times 10$ 
clusters have not revealed any clear sign of its stability either.
\end{thebibliography}
\end{document}